\documentclass[prl,twocolumn,nofootinbib]{revtex4}
\usepackage{graphicx}
\usepackage{latexsym}

\tighten
\begin{document}

\title{Domain Wall in the Linear Sigma Model\footnote{
Supported in part by the National Natural Science Foundation of
China under Grant No 10275070.}}
\author{MAO Hong$^{1,3}$\footnote{Email: maohong@mail.ihep.ac.cn}, LI Yunde$^{1,2}$, HUANG Tao$^1$}
\address{1. Institute of High Energy Physics, Chinese Academy of Sciences, Beijing
100039, China \\
2.Physics Department, East China Normal University, Shanghai,200062, China\\
3.Graduate School of the Chinese Academy of Sciences, Beijing
100039, China}
\begin{abstract}

We discuss the role of the axial $U(1)_A$ symmetry in the chiral
phase transition using the $U(N_f)_R\times U(N_f)_L$ linear sigma
model with two massless quark flavors. We expect that above a
certain temperature the axial $U(1)_A$ symmetry will be
effectively restored as well as $SU(N_f)_R\times SU(N_f)_L$. Then
we can construct a string-like static solution of the $\eta$
string and a kink-like classical solution of the domain wall
during the chiral phase transition. We give out the possible
signals for detecting the domain wall in ultrarelativistic
heavy-ion collisions.
\newline PACS number(s): 25.75.-q, 12.39.Fe, 98.80.Cq
\end{abstract}

\maketitle

Exploring the phase structure of quantum chromodynamics(QCD) is
one of the primary goals of ultrarelativistic heavy-ion
collisions. It is generally believed that at sufficiently high
temperature there should be a transition from ordinary hadronic
matter to a chirally symmetric plasma of quark and gluons \cite
{Pisarski84}. The order parameter for this phase transition is the
quark-antiquark condensate. At temperature of about 150 MeV,
Lattice QCD calculations indicate that this symmetry is restored
\cite{Peikert99}. The order of the phase transition seems to
depend on the mass of the non-strange u and d quarks, $m_u\approx
m_d$, and the mass of the strange quark $m_s$, and at the
temperature on the order of 150 MeV, heavier quark flavors do not
play an essential role.

For $N_f$ massless quark flavors, the QCD Lagrangian possesses a
chiral $U(N_f)_R\times U(N_f)_L=SU(N_f)_R\times SU(N_f)_L\times
U(1)_V\times U(1)_A$ symmetry, here $V=R+L$, while $A=R-L$.
However this symmetry does not appear in the low energy particle
spectrum, it is spontaneously broken to the diagonal $SU(N_f)_V$
group of vector transformation by a non-vanishing expectation
value of the quark-antiquark condensate, $\langle
\overline{q}_Rq_L\rangle\neq 0$. This process involves $N^2_f $
Goldstone bosons which dominate the low-energy dynamics of the
theory. The $U(1)_V$ symmetry is always respected and thus plays
no role in the symmetry breaking pattern considered in the
following discussion. The axial $U(1)_A$ symmetry is broken to
$Z(N_f)_A$ by a non-vanishing topological susceptibility
\cite{tHooft76}. Consequently, one of the $N^2_f$ Goldstone bosons
becomes massive, leaving $N^2_f-1$ Goldstone bosons. The
$SU(N_f)_R\times SU(N_f)_L\times U(1)_A$ group is also explicitly
broken by the effects of nonzero quark masses.

As the temperature or the density of matter increases, it is
expected that the instanton effects will rapidly disappear, the
$U(1)_A$ symmetry may also be effectively restored in addition to
$SU(N_f)_R\times SU(N_f)_L$. Since the chiral condensate $\langle
\overline{q}_Rq_L\rangle\neq 0$ also breaks the $U(1)_A$ axial
symmetry, there are only two possibilities: either the $U(1)_A$
symmetry is restored at a temperature much greater than the
$SU(N_f)_R\times SU(N_f)_L$ symmetry or the two symmetries are
restored at (approximately) the same temperature. Recent lattice
gauge theory computations have demonstrated a rapid dropping of
the topological susceptibility around the chiral phase transition,
seemingly suggesting that the simultaneous restoration exists
\cite{Alles00}, this is also supported by the
random matrix models\cite{Janik99}. On the other hand, the fate of the $%
U(1)_A$ anomaly in nature is not completely clear since instanton
liquid model calculations indicate that the topological
susceptibility is essentially unchanged at $T_c$ \cite{Schafer96},
also Lattice results obtained from the $SU(3)$ pure gauge theory
show that the topological susceptibility is approximately constant
up to the critical temperature $T_c$, it has a sharp decrease
above the transition, but it remains to be different from zero up
to $\sim1.2T_c$ \cite{Alles97}. Additionally, other lattice
computations which measure the chiral susceptibility find that the
$U(1)_A$ symmetry restoration is at or below the $15\%$ level
\cite{Bernard97} \cite {Chandrasekharan99}.

Recently, the issue of finding signals for the restoration of
chiral symmetry in ultrarelativistic heavy-ion collisions has
received considerable attention. For example, the signals for the
restoration of the $SU(2)$ chiral symmetry associated with the
$\sigma$ meson have been proposed in Refs.\cite
{Song97}\cite{Chiku98}. In particular, signals for detecting the
effective restoration of the $U(1)_A$ chiral symmetry in
ultrarelativistic heavy-ion collisions have been invoked
in\cite{Kharzeev98}\cite{Schaffner-Bielich00}\cite{Marchi03}.

On the other hand, in QCD, spontaneous symmetry breaking
$U(N_f)_R\times U(N_f)_L\rightarrow U(N_f)_V$ in the chiral limit
allows for existence of topological string defects, the formation
and evolution of such defects and their possible observable
effects in ultrarelativistic heavy-ion collisions as well as in
the early universe transition have been invoked in Refs.
\cite{Zhang98}\cite{Brandenberger99}\cite{Balachandran02}. In this
letter, we study the effects from effective restoration of the
$U(1)_A$ symmetry by using the $U(N_f)_R\times U(N_f)_L$ linear
sigma model with two massless flavors.

The Lagrangian of the $U(N_f)_R \times U(N_f)_L$ linear sigma
model is given by \cite{Roder03}
\begin{eqnarray}  \label{lag1}
{\mathcal L}(\Phi)&=&{\text{Tr}}(\partial_{\mu}\Phi^{+}
\partial^{\mu}\Phi-m^2
\Phi^{+} \Phi)-\lambda_{1}[{\text{Tr}}(\Phi^{+}\Phi)]^2  \nonumber \\
&& -\lambda_{2}{\text{Tr}}(\Phi^{+}\Phi)^2 +c[{\text{det}}(\Phi)+{\text{det}}%
(\Phi^{+})]  \nonumber \\
&& +{\text{Tr}}[H(\Phi+\Phi^{+})].
\end{eqnarray}
Where $\Phi$ is a complex $N_f \times N_f$ matrix parametrizing
the scalar and pseudoscalar mesons,
\begin{equation}
\Phi=T_a \phi_a=T_a(\sigma_a+i\pi_a),
\end{equation}
with $\sigma_a$ being the scalar ($J^p=0^+$) fields and $\pi_a$
being the pseudoscalar ($J^p=0^-$) fields. The $N_f \times N_f$
matrix $H$ breaks the symmetry explicitly and is is chosen as
\begin{equation}
H=T_a h_a,
\end{equation}
where $h_a$ are external fields, $a=0,1,\cdots,N_f^2-1$ and $T_a,
a\neq 0$ are a basis of generators for the $SU(N_f)$ Lie algebra.
$T_0={\mathbf{1}}$ is the generator for the $U(1)_A$ Lie algebra.

In the above model, the determinant terms correspond to the
$U(1)_A$ anomaly, as shown by 't Hooft \cite{tHooft76}, they arise
from instantons. These terms are invariant under $SU(N_f)_R \times
SU(N_f)_L \cong SU(N_f)_V \times SU(N_f)_A$, but break the
$U(1)_A$ symmetry of the Lagrangian explicitly. The last term in
Eq.(\ref{lag1}) which is due to nonzero quark masses breaks the
axial and possibly the $SU(N_f)_ V$ vector symmetry explicitly.

When $h_a=0, c=0$, for $m^2<0$ the global $SU(N_f)_V \times
U(N_f)_A$ symmetry is broken to $SU(N_f)_V$, and $\langle \Phi
\rangle$ develops a
non-vanishing vacuum expectation value, $\langle \Phi \rangle=T_0 \overline{%
\sigma}_0$. Spontaneously breaking $U(N_f)_A$ beads to $N_f^2$
Goldstone bosons which form a pseudoscalar, $N_f^2$ dimensional
multiplet. However when $h_a=0$, and $c\neq0$, the $U(1)_A$ is
further broken to $Z(N_f)$ by the axial anomaly, and $SU(N_f)_V
\times SU(N_f)_A$ is still the symmetry of the Lagrangian. A
nonvanishing $\langle \Phi \rangle$ spontaneously breaks the
symmetry to $SU(N_f)_V$, with the appearance of $N_f^2-1$
Goldstone bosons which form a pseudoscalar, $N_f^2-1$ dimensional
multiplet. The $N_f^2$th pseudoscalar meson is no longer massless,
because the $U(1)_A$ symmetry is already explicitly broken, e.g
for $N_f=2$, the $\eta$ meson is massive compared to other
pseudoscalar mesons. All these symmetry are in addition explicitly
broken by non-zero quark masses making all the Goldstone bosons
massive.

In the present study, since we only concentrate on the effects of
the effective restoration of the $U(1)_A$ symmetry, we can ignore
the possible effects of the restoration of $SU(2)_R \times
SU(2)_L$, this implies that we can forget $\pi$ and $a_0$ fields,
keeping only the $\sigma$ and $\eta$ mesons which are usually
specified by the $U(1)_A$ phase. With this restriction on $\Phi$,
the effective Lagrangian we adopt here is
\begin{eqnarray}  \label{lag2}
{\mathcal L}(\Phi)&=&{\text{Tr}}(\partial_{\mu}\Phi^{+}
\partial^{\mu}\Phi-m^2
\Phi^{+} \Phi)-\lambda_{1}[{\text{Tr}}(\Phi^{+}\Phi)]^2  \nonumber \\
&& -\lambda_{2}{\text{Tr}}(\Phi^{+}\Phi)^2 +c[{\text{det}}(\Phi)+{\text{det}}%
(\Phi^{+})],
\end{eqnarray}
where $\Phi=\frac{1}{2}(\sigma+i\eta){\mathbf{1}}$. In the
following, we demonstrate that both a static string-like solution
of the $\eta$ string and a static kink-like solution of the domain
wall are expected to be produced during the chiral phase
transition\footnote{For simplicity we consider here the
configurations which are specified by the $U(1)_A$ phase only. In
considering non-abelian phases, there is another class of
topological defects known as non-abelian
strings\cite{Balachandran02}, the pion strings, which can also
exist during the chiral phase transition\cite{Zhang98}.}.

The $\eta$ string is a static configuration of the Lagrangian of
Eq.(\ref {lag2}) with $c=0$. In this case, during chiral symmetry
breaking, the field $\sigma$ takes on a nonvanishing expectation
value, which breaks $U(2)_R\times U(2)_L$ down to $U(2)_V$. This
results in a massive $\sigma$ and four massless Goldstone bosons.

In our discussion of the $\eta$ string and domain walls it is
convenient to define the new fields
\begin{equation}  \label{phi}
\phi=\frac{\sigma+i \eta}{\sqrt{2}}.
\end{equation}
The linear sigma model in Eq.(\ref{lag2}) with $c=0$ now can be
rewritten as
\begin{equation}  \label{lag3}
{\mathcal{L}}=(\partial_{\mu}\phi)^*(\partial^{\mu}\phi)-\lambda(\phi^*
\phi-\frac{v^2}{2})^2,
\end{equation}
where $v^2=\frac{-m^2}{\lambda}$ and
$\lambda=\lambda_1+\frac{\lambda_2}{2}$. For static
configurations, the energy functional corresponding to the above
Lagrangian is
\begin{equation}  \label{energy1}
E=\int
d^3x[\nabla\phi^*\nabla\phi+\lambda(\phi^*\phi-\frac{v^2}{2})^2],
\end{equation}
and the time independent equation of motion is
\begin{eqnarray}
\nabla^2\phi=2\lambda(\phi^{*}\phi-\frac{v^2}{2})\phi.
\end{eqnarray}
The $\eta$ string solution extremising the energy functional of
Eq.(\ref {energy1}) is given in Refs.
\cite{Zhang98}\cite{Vilenkin00}.
\begin{equation}
\phi =\frac{v }{\sqrt{2}}\rho(r)\exp (i\theta ),
\end{equation}
where $\rho(r)=[1-\exp (-\mu r)]$, the coordinates rand are polar
coordinates in the $x-y$ plane, the $\eta$ string is assumed to
lie along the $z$ axis and $\mu^2=\lambda\frac{89}{144}v^2$. The
energy per unit length of the string is
\begin{equation}
E=[0.75+\log (\mu R)]\pi v^2.
\end{equation}
For global symmetry in general the energy density of the string
solution is logarithmically divergent, $R$ is introduced as a
cutoff which is taken to be $O( \mathrm{fm})$ in the following
numerical calculation.

In the case of $c\neq 0$, during chiral symmetry breaking, the
field $\sigma$ takes on a nonvanishing expectation value, which
breaks $SU(2)_R\times SU(2)_L$ down to $SU(2)_V$. This results in
a massive $\sigma$ and three massless Goldstone bosons, in the
same time the $\eta$ meson is massive compared to other
pseudoscalar mesons. Then the determinant term in Eq.(\ref {lag2})
can not be simplistically neglected during the chiral phase
transition in nature, so that one of the appropriate description
is no longer one of the $\eta$ strings, but one of domain walls.
Then in the following discussion we only consider the possible
effects of domain walls and ignore the possible effects of the
$\eta$ string in the ultrarelativistic heavy-ion collisions. With
the defination of new fields in Eq.(\ref{phi}), the Lagrangian of
Eq.(\ref{lag2}) can be simplistically expressed as
\begin{equation}  \label{lag4}
{\mathcal{L}}=(\partial_{\mu}\phi)^*(\partial^{\mu}\phi)-m^2\phi^*
\phi+c \text{Re}(\phi^2)-\lambda(\phi^*\phi)^2.
\end{equation}
After defining $c=\alpha m^2$, the potential takes the form
\begin{equation}
V(\phi)=\lambda(\phi^*\phi)^2-m^2(\alpha \text{Re}(\phi^2)-\phi^*
\phi).
\end{equation}
The limit $\alpha\rightarrow\infty$ corresponds to the maximum
explicit $U(1)_A$ symmetry breaking. In this limit, for realistic
values of the $\sigma$ meson and the $\pi$ meson mass(i.e.,
$m^2-c=constant$), the $\eta$ and $a_0$ mesons become infinitely
heavy and are thus removed from the spectrum of physics
excitations, and $U(2)_R\times U(2)_L$ is identical to the $O(4)$
model, there has no $\eta$ strings and domain walls. For the
chiral symmetry spontaneously breaking to occur, we always require
$\alpha>1$. In the following numerical calculation, we take
$c=(386.79MeV)^2$, for other parameters we have
$\lambda_1=-31.51$, $\lambda_2=82.77$ and $m^2=(263.83MeV)^2$
corresponding to $m_{\sigma}=400MeV$ and
$m_{\eta}=547MeV$\cite{Roder03}.

For static configuration in Eq.(\ref{lag4}), the energy functional
is given by
\begin{equation}  \label{energy2}
E=\int
d^3x[\nabla\phi^*\nabla\phi+\lambda(\phi^*\phi)+m^2(\phi^*\phi)-\alpha m^2%
\text{Re}(\phi^2)].
\end{equation}
The corresponding equation of motion for the field $\phi$ is
\begin{equation}
\nabla^2\phi+m^2(\alpha \phi^*-\phi)-2\lambda|\phi^2|\phi=0,
\end{equation}
which accepts the static symmetric kink
solution\cite{Vilenkin00}\cite {Axenides97}
\begin{eqnarray}
\sigma &=& m\sqrt{\frac{(\alpha-1)}{\lambda}}\text{tanh}[\sqrt{\frac{%
(\alpha-1)}{2}}m x], \\
\eta &=& 0.
\end{eqnarray}
The thickness of this wall is approximately
\begin{equation}
\delta \sim (m\sqrt{\alpha-1})^{-1} \simeq 0.7 fm,
\end{equation}
and the mass per unit area of the walls is
\begin{equation}  \label{energy3}
\omega=\frac{2\sqrt{2}m^3}{3\lambda}(\alpha-1)^{\frac{3}{2}}\simeq
(129.273 MeV)^3.
\end{equation}
The stability becomes a consequence of a topological conservation
law. The
topological current from which this law is derived $j^\mu=\epsilon^{\mu\nu}%
\partial_\nu \phi$, the associated charge of a configuration is $N=\int
dxj^0=\phi|_{x=+\infty}-\phi|_{x=-\infty}$, the presence of a kink
with $\phi $ in different vacuum at $x=\pm\infty$, gives rise to a
non-zero charge $N$ and consequently indicates the stability of
the configuration. Moreover, the form of the potential implies
that the symmetric wall solution (within the domain wall the
$\eta=0$) is dynamically stable. We consider infinitesimal
perturbations of the field $\eta$ and check if the variation in
the energy is positive or negative. Discarding terms of cubic and
higher orders in $\eta $, we find
\begin{eqnarray}  \label{energy4}
E=E_{(domain wall)}+\delta E,
\end{eqnarray}
where
\begin{eqnarray}
\delta E=\int d^3x[\frac{1}{2}\vec{\nabla}\eta \vec{\nabla}\eta+\frac{1}{2}%
(\alpha+1)m^2\eta^2+\frac{\lambda}{4}\sigma^2 \eta^2].
\end{eqnarray}
From the above equation, the term $\delta E$ in
Eq.(\ref{energy4})is always positive, therefore, the domain walls
of the Lagrangian (\ref{lag2})is topologically stable and
dynamically stable.

In the Sine-Gorden model, the kink solutions are absolutely stable
and such a stable domain wall will immediately rule out by the
cosmological constraint in general. In our case, the domain wall
is only metastable in full theory since there are other dynamical
fields corresponding to the remaining $SU(2)$ generators (such as
$\pi$ and $a_0$ fields). However, one can show that these
dynamical fields do not contribute to the domain wall background
but simply remain in their vacuum states. Their fluctuations
affect the overall energy density, but do not affect the
properties of the domain wall such as the surface tension and so
we can neglect their effects\cite{Forbes01}. Then the domain wall
can still be taken as classically stable object, and therefore, it
decays through the quantum tunnelling process with exponentially
large lifetime which is longer than any other time scales existing
in the ultrarelativistic heavy ion collisions\cite{Shuryak02}.
Then all the pions which are eventually emitted from such an
object will be completely incoherent with the rest of pions.

In the ultrarelativistic heavy-ion collisions, domain walls are
expected to be produced during the chiral phase transition. If a
bubble wall is produced\cite{Shuryak02}, it exists for some
lifetime and then decays into its underling fields, the $\sigma $
fields. We make the assumption that the size of the bubble wall
should be around the size of the QGP formed at ultrarelativistic
heavy-ion collisions. The experimental observation of the domain
wall bubbles can be carried out by using the Hanbury-Brown-Twiss
(HBT) intensity interferometry of
pions\cite{Zhang2001}\cite{Wu02}. As pointed by Shuryak and
Zhitnitsky in Ref.\cite{Shuryak02} if a bubble exists for enough
long time($\sim $5 fm) and then decays the bubble can be taken as
an long-lived object. Therefore the pions from the bubble lead to
the same effect of not producing an HBT peak in two-pion spectra
which is just as that of the long-lived hadronic resonances. To
see this, an effective intercept parameter, $\lambda _{eff}$, is
introduced~~in Bose-Einstein correlation function\cite{Vance98}
\begin{equation}
C_2(k,K)=\frac{N_2(p_1,p_2)}{N_1(p_1)N_2(p_2)}=1+\lambda
_{eff}(p)R_c(k,K),
\end{equation}
where the effective intercept parameter and the correlator are
given by
\begin{equation}
\lambda _{eff}(p)=\left[ \frac{N_c(p)}{N_c(p)+N_h(p)}\right] ^2
\end{equation}
and
\begin{equation}
R_c(k,K)=\frac{\left| \widetilde{S_c}(k,K)\right| ^2}{\left| \widetilde{S}%
_c(k=0,K=p)\right| ^2},
\end{equation}
where $k=(p_1-p_2)$ , ~$K=(p_1+p_2)/2$ , $N_c(N_h)$ is the
one-particle invariant momentum distribution of the
``core''$~($~and ``halo'' )$~$~decayed pions
respectively.$~~\widetilde{S}_c$ is the Fourier transform of the
one-boson emission function. The produced bubbles would given an
additional factor $\beta$ to the effective intercept.

\begin{eqnarray}
\lambda _{eff}' &=& \left[
\frac{N_c}{N_c+N_h+N_{domain-wall}}\right] ^2 \nonumber
\\&\approx&
\left[ 1-\frac{N_{domain-wall}}{N_c+N_h}\right] ^2\left[ \frac{N_c}{N_c+N_h}%
\right] ^2 \nonumber \\
&=& \beta\lambda _{eff},
\end{eqnarray}
where $N_{domain-wall}$ is the number of pions from the decay of
domain wall bubbles. In RHIC Pb-Pb collisions$~$if we take the
radius of QGP phase as the domain wall bubble radius $R\sim
6fm$\cite{Zhang01} , then the domain wall bubble energy is about
$E_{domainwall}\simeq 4\pi R^2\omega \approx 25GeV$, . If all the
energy accumulated in the wall will lead to the production of the
$\sigma $ mesons(which will result in additional $\sim 60$ mesons
per event) one should expect a 40 $\pi ^{+}($or $\pi ^{-}$) to be
produced from the bubble wall in the central rapidity region. At
RHIC energy the total number of pions is about 1500, so the factor
is about $\beta \sim 0.85$ . In the case of LHC Pb-Pb collisions
the QGP radius is about 10 fm\cite {Zhang01}, this gives out
$\beta \sim 0.7-0.8$. Thus we can use pion interferometry as a
sensitive tool to detect this possible increase of the $\sigma $
production in ultrarelativistic heavy-ion collisions.

In summary, we have discussed the possible effects of the
restoration of the axial $U(1)_A$ symmetry during the chiral phase
transition by using the $U(N_f)_R\times U(N_f)_L$ linear sigma
model with two massless quark flavors. It is emphasized that if
the axial $U(1)_A$ symmetry is to be restored above the certain
temperature, it is the domain wall rather than the $\eta$ string
that is expected to be produced and has a long lifetime then the
time scale existing in the ultrarelativistic heavy-ion collisions.
These domain walls will decay into the $\sigma$ mesons, and the
increase of the $\sigma$ mesons can be viewed as a signal of
restoration of the axial $U(1)_A$ symmetry in ultrarelativistic
heavy-ion collisions.

\begin{acknowledgments}
The authors wish to thank Michiyasu Nagasawa, Nicholas Petropoulos
and Xinmin Zhang for useful discussions and correspondence.
\end{acknowledgments}

\end{document}